# Deep residual learning in CT physics: scatter correction for spectral CT

Shiyu Xu*, Peter Prinsen, Jens Wiegert and Ravindra Manjeshwar

*Abstract*– Recently, spectral CT has been drawing a lot of attention in a variety of clinical applications primarily due to its capability of providing quantitative information about material properties. The quantitative integrity of the reconstructed data depends on the accuracy of the data corrections applied to the measurements. Scatter correction is a particularly sensitive correction in spectral CT as it depends on system effects as well as the object being imaged and any residual scatter is amplified during the non-linear material decomposition. An accurate way of removing scatter is subtracting the scatter estimated by Monte Carlo simulation. However, to get sufficiently good scatter estimates, extremely large numbers of photons are required, which may lead to unexpectedly high computational costs. Other approaches model scatter as a convolution operation using kernels derived using empirical methods. These techniques have been found to be insufficient in spectral CT due to their inability to sufficiently capture object dependence. In this work, we develop a deep residual learning framework to address both issues of computation simplicity and object dependency. A deep convolution neural network is trained to determine the scatter distribution from the projection content in training sets. In test cases of a digital anthropomorphic phantom and real water phantom, we demonstrate that with much lower computing costs, the proposed network provides sufficiently accurate scatter estimation.

## I. INTRODUCTION

SPECTRAL CT provides additional information, on top of conventional CT, enabling more accurate diagnostics [1]. The spectral results are divided into several categories: virtual monochromatic images enabling reduced image artifacts and enhancement of contrast; material separation such as water - Iodine, Calcium - Iodine, Calcium – uric-acid; material characterization such as effective atomic number. However, there are several limitations of current spectral systems, such as inaccurate material separation and unstable image quality (IQ) which are driven by small inaccuracies in scanner calibration and data corrections. The residual errors, which are not so impactful in conventional CT, are amplified during the non-linear material decomposition in the Spectral CT image chain [2]. Scatter correction is a particularly sensitive correction in spectral CT as it depends on system effects as well as the object being imaged. The subtle residual scatter does matter regardless of whether projection-based decomposition or image-based decomposition is performed.

There are many scatter compensation techniques. An anti-scatter grid could dramatically reduce the large angle scatter signal but could also decrease primary signal and insufficiently filter out small angle scatter and multi-scatter. The most efficient way to estimate scatter is a model-based method that defines a system and scatter model and runs Monte Carlo simulations to calculate the scatter background[3], [4]. However, to get accurate simulation results, large numbers of photons need to be simulated, which leads to long computation times. Other approaches model scatter as a convolution operation using kernels derived using empirical methods. These techniques have been found to be insufficient in spectral CT due to their inability to sufficiently capture object dependence[5].

Recently, deep learning approaches have achieved tremendous success in many fields. In medical imaging for example, there have been extensive research activities applying deep learning. However, most of these efforts are focused on image-based diagnostics[6]. There are a few investigations related to CT imaging[7], [8]. In this work we bring this promising technique into imaging physics. The main goal is to develop a novel deep convolutional neural network (CNN) architecture for scatter correction that approaches the quality of model-based methods with much lower computing costs.

The proposed deep residual learning is based on our conjecture that the scatter distribution of a pencil beam can be determined from its corresponding air-normalized raw signal (primary plus scatter) and the signal in the surrounding area. Therefore, to train the network, the input is a projection with air-normalized raw signal and its label is the corresponding scatter signal, which is calculated in a Monte Carlo simulation. The idea here takes advantage of a residual learning framework where the network learns the small offset scatter signal from the large raw signal.

## II. METHODS AND EXPERIMENTS

### A. CNN architectures

The very deep ConvNets known as VGG [9] were used for large-scale visual recognition. We change VGG to make it suitable for scatter estimation, and set the depth of the network based on the required receptive field. For model learning, we adopt the residual learning formulation, and incorporate it with batch normalization for fast training and improved performance.

Following the methods in [9], [10], we set the size of the convolution filter to 3x3 but remove all pooling layers and

---

fully connected layers. We downsample the dimension of each projection to 84 radial detectors and 8 slices. The depth of the network d and kernel size are designed based on the required receptive area of the network, which is expected to be the entire input projection due to the characteristics of scatter distribution.

We choose the depth of the network d = 22. The first layer consists of 64 filters of size 3x3, layers 2-20 each consists of 64 filters of size 3x3x64 with dilate factor 2[11], and the last layer consists of a single filter of size 3x3x64. Except for the first and last layer, each convolution layer is followed by a batch normalization, which is included to speed up training as well as boost performance[10], and rectified linear units (ReLU), which are used to introduce nonlinearity. Simple zero padding is performed in each convolution layer to maintain the data dimensions.

The input of our CNN (r) is air-normalized primary signal (p) with additional scatter (s): r = p + s. We adopt the residual learning formulation to train a residual mapping $T(r) \sim s$, from which we determine the desired signal $p = r - T(r)$. The CNN parameters are estimated by minimizing the following loss function:

$$L(w) = \sum_j (\left\| T(r;w)_j - s_j \right\|_2^2 + \lambda_1 \left\| \nabla T(r;w)_j \right\|_1) + \lambda_2 \sum_k \left\| w_k \right\|_2^2$$

Here w is the set of all convolutional kernels of all layers and k=1,…,22 denotes the layer index. The regularization terms encourage a smoothed scatter signal by total variation constraint [12] and small network weights. We used the regularization parameters $\lambda_1 = \lambda_2 = 10^{-3}$. Here $\{(r_j, s_j)\}_{j=1}^N$ represents N training pairs of scattered raw signal and scatter-only signal, where j is the index of training unit. The training sets are obtained from MC simulations.

The minimization of the cost function was performed using the backpropagation with Adam optimizer[13], where an initial learning rate was set to 0.001, and the learning rate was gradually decreased to $10^{-6}$. Mini-batches of size 50 were used, indicating that 50 randomly chosen sets of data were used as a batch. The method was implemented in MATLAB using MatConvNet[14].

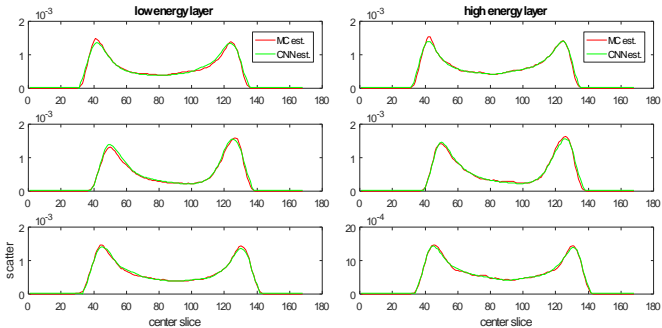

Fig. 1. Scatter profiles from the CNN and from MC (as ground truth) at the center slice for three different views (from top to bottom). X axis is downsampled detector line; y axis is normalized scatter intensity.

### B. Training sets

A model-based Monte Carlo simulation was employed to generate the training sets. The geometry setup and parameters of the simulations were chosen to mimic the characteristics of the Philips IQon Spectral CT, which has a dual-layer detector. The simulations were performed with a tube voltage of 120 kVp. Three phantoms were used: a 30 cm diameter cylindrical water phantom, and digital representations of an anthropomorphic liver phantom and an obese phantom. For each simulated scan, the primary signal was determined analytically using 600 (obese and liver) or 2400 (water) projections over a full rotation of 360 degrees. Scatter data was collected by simulating 60 projections over a full rotation with $5x10^8$ (water and liver) or $4.5x10^9$ (obese) photons per projection. Due to the expensive computation, a limited number of 8 representative scans were used to generate

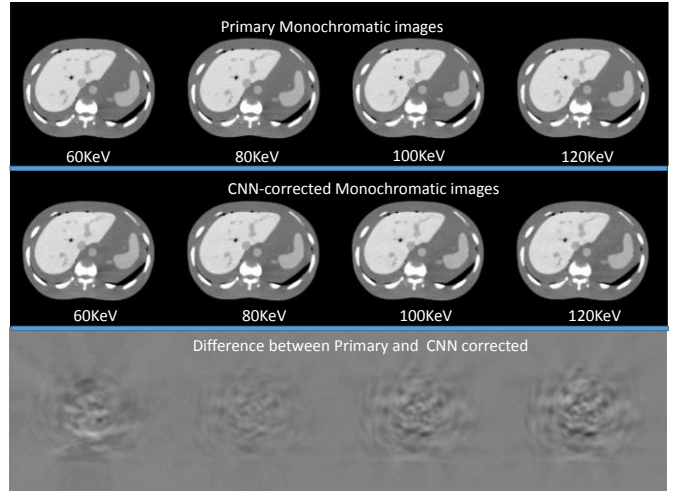

Fig. 2. CNN-corrected monochromatic images ([-50 50]) at 60keV, 80keV, 100keV and 120keV versus Primary monochromatic image ([-50 50])) and their difference [-10 10].

training sets: 1 for the water phantoms and 3 for the obese phantom, and 4 for the liver phantom at different shifts of the phantom in cranio-caudal direction. Data was augmented by simply flipping along detector line dimension. The network was trained separately for the low energy layer and the high energy one. The total training time for each layer was about 8

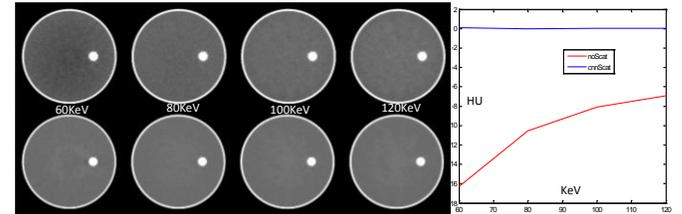

Fig. 3. CNN-corrected monochromatic images ([-50 50]) at 60keV, 80keV, 100keV and 120keV on the bottom versus monochromatic without scatter correction ([-50 50])) on the top. The dispersion curves on the right are drawn by determining the average HU value in the water phantom for each keV, which is supposed to be flat at zero for water across keVs.

hours on a Dell T7600 workstation with a Titan X GPU.

### C. Testing sets

For testing the same liver phantom as above was used but at a cranio-caudal shift that was not used for training. A 30cm water cylinder and liver region of an anthropomorphic phantom were used as well but this time scanned on a Philips IQon system, not simulated. We estimated scatter by applying the well-trained network to the low and high-energy signals separately. Scatter correction was performed by simply subtracting the corresponding estimation from the raw data of each layer. The corrected high and low data were then further decomposed into energy independent basis functions, such as Iodine and Water. These were reconstructed into basis functions in image space. Monochromatic images were constructed as linear combinations of basis functions.

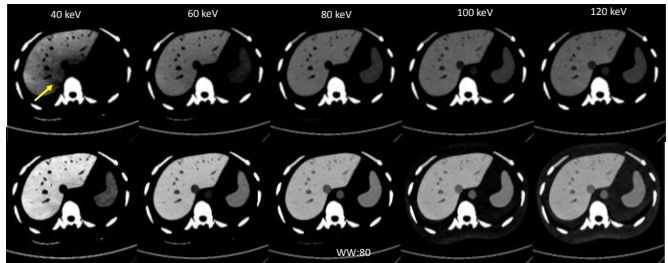

Fig. 4. CNN-corrected monochromatic images ([-40 40]) at 40keV, 60keV, 80keV, 100keV and 120keV on the bottom versus monochromatic without scatter correction ([-40 40])) on the top.

## III. RESULTS

For the simulated data, scatter profiles from the CNN at the center slice of three different views are plotted in Fig. 1, and compared to those from MC simulation (as ground truth) for both detector layers. One can see that CNN estimated scatter is matching very well to MC simulated scatter. After applying the CNN-based scatter correction, monochromatic images at 60keV, 80keV, 100keV and 120keV are reconstructed and compared to the monochromatic images obtained from reconstructing the primary signal, see Fig. 2. The difference shown in the third row represents the residual scatter after CNN based correction, which shows no structured bias or shading from the CNN based method.

For the real water phantom data, we present monochromatic images from 60keV to 120keV, draw dispersion curves, and compare the results with results without scatter correction, see Fig. 3. The flat dispersion curve indicates that the CNN based scatter correction could improve the accuracy of quantitative measurement. In the real liver phantom experiment, the monochromatic images from 40keV to 120keV are reconstructed and compared to images without scatter correction, see Fig.4. Beyond the CT number difference between the two rows, one can see severe shading artifacts appear along edge of the liver in the 40keV image without scatter correction, which is caused by a sharp change in the scatter distribution along the arrow direction, while the artifacts are corrected in the CNN based method.

In addition, the computing time to estimate scatter for each projection is about 10 ms on a single GPU enabled workstation.

## IV. CONCLUSION AND DISCUSSION

We developed a deep residual learning framework for scatter correction on spectral CT systems. We demonstrated that the proposed method provides similar performance to Monte Carlo simulation-based scatter correction but with a much shorter computing time. The efficacy of the proposed method was also illustrated on real phantom data. We believe that this suggests a new innovative framework for the CT physics field.

ACKNOWLEDGMENT

We thank Dr. Boaz Keren-Zur and Dr. Yoad Yagil for their valuable comments to improve the work.